\documentclass[journal=jacsat,manuscript=article]{achemso}
\usepackage{chemformula} % Formula subscripts using \ch{}
\usepackage[T1]{fontenc} % Use modern font encodings
\usepackage{amsmath,amssymb}
\usepackage{xcolor,soul}

\usepackage{atbegshi}
\AtBeginDocument{\AtBeginShipoutNext{\AtBeginShipoutDiscard}}

\author{Aaron L. Sharpe}
\affiliation[APPPhys]
{Department of Applied Physics, Stanford University, 348 Via Pueblo Mall, Stanford, CA 94305, USA}
\alsoaffiliation[SIMES]
{Stanford Institute for Materials and Energy Science, SLAC National Accelerator Laboratory, 2575 Sand Hill Road, Menlo Park, California 94025, USA}
\altaffiliation{Present address: Quantum and Electronic Materials Department, Sandia National Laboratories, Livermore CA 94550, USA}
\email{asharpe@sandia.gov}
\author{Eli J. Fox}
\affiliation[PHYS]
{Department of Physics, Stanford University, 382 Via Pueblo Mall, Stanford, CA 94305, USA}
\alsoaffiliation[SIMES]
{Stanford Institute for Materials and Energy Science, SLAC National Accelerator Laboratory, 2575 Sand Hill Road, Menlo Park, California 94025, USA}
\author{Arthur W. Barnard}
\affiliation[PHYS]
{Department of Physics, Stanford University, 382 Via Pueblo Mall, Stanford, CA 94305, USA}
\alsoaffiliation[SIMES]
{Stanford Institute for Materials and Energy Science, SLAC National Accelerator Laboratory, 2575 Sand Hill Road, Menlo Park, California 94025, USA}
\alsoaffiliation[UW]
{Department of Physics and Department of Materials Science and Engineering, University of Washington, 302 Roberts Hall, Seattle, Washington 98195, USA}
\author{Joe Finney}
\affiliation[PHYS]
{Department of Physics, Stanford University, 382 Via Pueblo Mall, Stanford, CA 94305, USA}
\alsoaffiliation[SIMES]
{Stanford Institute for Materials and Energy Science, SLAC National Accelerator Laboratory, 2575 Sand Hill Road, Menlo Park, California 94025, USA}
\author{Kenji Watanabe}
\affiliation[NIMS0]{Research Center for Functional Materials,
National Institute for Materials Science, 1-1 Namiki, Tsukuba 305-0044, Japan}
\author {Takashi Taniguchi}
\affiliation[NIMS1]
{International Center for Materials Nanoarchitectonics,
National Institute for Materials Science,  1-1 Namiki, Tsukuba 305-0044, Japan}
\author{M. A. Kastner}
\affiliation[PHYS]
{Department of Physics, Stanford University, 382 Via Pueblo Mall, Stanford, CA 94305, USA}
\alsoaffiliation[SIMES]
{Stanford Institute for Materials and Energy Science, SLAC National Accelerator Laboratory, 2575 Sand Hill Road, Menlo Park, California 94025, USA}
\alsoaffiliation[MIT]
{Department of Physics, Massachusetts Institute of Technology, 77 Massachusetts Avenue, Cambridge, MA 02139, USA}
\author{David Goldhaber-Gordon}
\affiliation[PHYS]
{Department of Physics, Stanford University, 382 Via Pueblo Mall, Stanford, CA 94305, USA}
\alsoaffiliation[SIMES]
{Stanford Institute for Materials and Energy Science, SLAC National Accelerator Laboratory, 2575 Sand Hill Road, Menlo Park, California 94025, USA}
\email{goldhaber-gordon@stanford.edu}

\title{Evidence of orbital ferromagnetism in twisted bilayer graphene aligned to hexagonal boron nitride}


\begin{document}
\pagenumbering{arabic}

%%%%%%%%%%%%%%%%%%%%%%%%%%%%%%%%%%%%%%%%%%%%%%%%%%%%%%%%%%%%%%%%%%%%%
%% ABSTRACT
%%%%%%%%%%%%%%%%%%%%%%%%%%%%%%%%%%%%%%%%%%%%%%%%%%%%%%%%%%%%%%%%%%%%%

\begin{abstract}
    We have previously reported ferromagnetism evinced by a large hysteretic anomalous Hall effect in twisted bilayer graphene (tBLG)~\cite{Sharpe2019}. Subsequent measurements of a quantized Hall resistance and small longitudinal resistance confirmed that this magnetic state is a Chern insulator~\cite{Serlin2020}. Here we report that, when tilting the sample in an external magnetic field, the ferromagnetism is highly anisotropic. Because spin-orbit coupling is negligible in graphene~\cite{Sichau2019} such anisotropy is unlikely to come from spin, but rather favors theories in which the ferromagnetism is orbital. We know of no other case in which ferromagnetism has a purely orbital origin. For an applied in-plane field larger than $5\ \mathrm{T}$, the out-of-plane magnetization is destroyed, suggesting a transition to a new phase. 
\end{abstract}

%%%%%%%%%%%%%%%%%%%%%%%%%%%%%%%%%%%%%%%%%%%%%%%%%%%%%%%%%%%%%%%%%%%%%
%% Introduction
%%%%%%%%%%%%%%%%%%%%%%%%%%%%%%%%%%%%%%%%%%%%%%%%%%%%%%%%%%%%%%%%%%%%%

The possibility of flat bands in Van der Waals heterostructures, beginning with magic-angle twisted bilayer graphene (tBLG)~\cite{Bistritzer2011, Cao2018a, Cao2018b}, has drawn much experimental and theoretical attention. In such weakly dispersing bands the kinetic energy is reduced, allowing the electron-electron interactions to favor correlated states. Evidence of correlated behavior has been observed, for example, as the appearance of resistive states at fractional filling of bands in tBLG with an interlayer twist of $\approx 1.1^\circ$~\cite{Cao2018a, Yankowitz2019}, ABC-trilayer graphene/hBN moir\'e~\cite{Chen2018, Chen2019}, twisted bi-bilayer~\cite{Liu2020, Cao2019, Burg2019, Vries2020, MHe2020a}, monolayer-bilayer graphene~\cite{Polshyn2020, SChen2020, Shi2020}, and twisted $\mathrm{WSe_2}$~\cite{Wang2020}. One recently observed consequence of these correlations is ferromagnetism in tBLG in a narrow range of carrier densities around $3/4$ filling of the flat conduction band,~\cite{Sharpe2019, Serlin2020} where full filling corresponds to four electrons per moir\'e unit cell accounting for spin and valley degeneracies~\cite{Cao2016}. The magnetism was initially revealed by a hysteretic anomalous Hall effect as large as $10.4\ \mathrm{k\Omega}$~\cite{Sharpe2019}. Initial evidence of chiral edge states from nonlocal transport~\cite{Sharpe2019} indicated that the magnetic state could be an incipient Chern insulator. Subsequent measurements of similarly-configured tBLG revealed precise quantization of the Hall resistivity coincident with longitudinal resistivity as small as $10\Omega$~\cite{Serlin2020}, conclusively demonstrating it is possible to achieve a Chern insulator with Chern number $C=1$ at 3/4 filling. At optimal doping, this Chern insulator has a coercive field of tens of millitesla and survives up to $9\ \mathrm{K}$.\cite{Serlin2020}. Evidence of Chern insulators has also been predicted and observed in ABC-trilayer graphene/hBN moir\'e superlattice~\cite{Zhang2020, Chen2020} and monolayer-bilayer graphene heterostructures~\cite{Polshyn2020, Chen2020}.  

A Chern insulator requires nontrivial band topology and a gap between bands of different Chern numbers. tBLG samples are generally encapsulated in hexagonal boron nitride (hBN) flakes to protect from disorder and serve as dielectrics for electrostatic gating. A gap at the Dirac point that would favor forming a Chern insulator could be opened by aligning the tBLG crystal axis to that of one of the cladding hBN layers, breaking $C_2$, the in-plane two-fold rotation symmetry~\cite{Bultinck2020, Zhang2019b, Zhu2020, Liu2021, Song2015}, that could otherwise protect the Dirac crossings of bands (in conjunction with time reversal symmetry $T$).  Though self-consistent Hartree-Fock calculations show that in-plane two-fold rotation symmetry may be broken spontaneously~\cite{Xie2020}, both tBLG samples that exhibit ferromagnetism appear to have hBN aligned to the tBLG~\cite{Sharpe2019, Serlin2020}, whereas such alignment has typically been intentionally avoided when fabricating other samples.

The precise nature of the ferromagnetic ground state is an open question. In ferromagnetic materials, exchange interactions break time reversal symmetry, favoring long-range order of electron spins. Though the motion of electrons can generate an orbital magnetic dipole moment, we are unaware of any magnetic material where the magnetism is dominated by the orbital magnetic moment independent of spin. However, for tBLG this is precisely the prediction of Refs.~\citenum{Bultinck2020, Zhang2019b, Zhu2020, Wu2020, Alavirad2019, Repellin2019b, Zhang2020, Liu2019, Liu2021}. Recent measurements using a superconducting quantum interference device found the magnetization to be approximately 2-4 Bohr magnetons per moir\'e unit cell~\cite{Tschirhart2020}. Those measurements were performed near the $3/4$ state, which corresponds to a single conduction-band hole per moir\'e unit cell. The measured magnetization significantly exceeds the expected 1 Bohr magneton per moir\'e unit cell for a single spin, suggesting that the magnetism has a strong orbital component. Here, we demonstrate by transport measurements in magnetic fields at angles ranging from normal to the plane of the sample to completely in the plane of the sample that the magnetism in tBLG is highly anisotropic. Given that spin-orbit coupling in graphene is weak~\cite{Sichau2019}, this observation implies that the magnetism is dominated by the orbital magnetic moment (expected to be highly anisotropic) rather than the isotropic spin. 

%%%%%%%%%%%%%%%%%%%%%%%%%%%%%%%%%%%%%%%%%%%%%%%%%%%%%%%%%%%%%%%%%%%%%
%% Fab and previous results
%%%%%%%%%%%%%%%%%%%%%%%%%%%%%%%%%%%%%%%%%%%%%%%%%%%%%%%%%%%%%%%%%%%%%

We used the ``tear-and-stack'' dry transfer method~\cite{Cao2016, Kim2016} and standard lithography techniques to fabricate a tBLG Hall bar device which was previously characterized in Ref.~\citenum{Sharpe2019}. The device was fabricated with both a silicon back gate and a Ti/Au top gate, allowing for independent control of the charge carrier density $n$ and perpendicular displacement field $D$ (see Methods)~\cite{Oostinga2008}. We measured the longitudinal resistance $R_{xx}$ and Hall resistance $R_{yx}$ using standard lock-in techniques with a 5-nA root mean square (RMS) AC bias current. The angle of the top graphene sheet relative to the top cladding hBN was measured by optical microscopy to be $1.0^\circ \pm 0.3^\circ$ clockwise (see the Supplemental Information). Features in electron transport apparently due to this alignment correspond to a hBN twist angle of $0.87^\circ \pm 0.6^\circ$ relative to the nearer graphene sheet. A rough measure of the twist angle between the sheets of graphene of $1.20^\circ \pm 0.1^\circ$ can be obtained from the superlattice density $n_s$ which corresponds to four electrons (or holes) per superlattice unit cell. $n_s$ is determined from the distance in gate voltage between the charge neutrality point (CNP) and the peak in resistance corresponding to $n_s$, multiplied by $dn/dV_{tg}$ extracted for top gate voltages $V_{tg}$ near the CNP~\cite{Hunt2013}. Typically, a more accurate measure of the twist angle can be obtained by fitting quantum oscillations. However, for this sample these features are not very sharp and yield a twist angle between the graphene sheets of $1.17^\circ \pm 0.05^\circ$. This twist angle corresponds to $\pm n_s = 3.18 \times 10^{12}\ \mathrm{cm^{-2}}$. This measure of the twist angle between the two graphene sheets is consistent with such a measurement performed on an atomic force microscopy image, which shows that the bottom graphene layer is rotated clockwise relative to the top graphene layer (see the Supplemental Image). This sample exhibits a magnetic state near $3/4$ filling.  Though the low-field ground state with strong and hysteretic anomalous Hall effect could in principle be metallic, we have identified it as an incipient Chern insulator~\cite{Sharpe2019}.

%%%%%%%%%%%%%%%%%%%%%%%%%%%%%%%%%%%%%%%%%%%%%%%%%%%%%%%%%%%%%%%%%%%%%
%% Results
%%%%%%%%%%%%%%%%%%%%%%%%%%%%%%%%%%%%%%%%%%%%%%%%%%%%%%%%%%%%%%%%%%%%%

By mounting the sample on a two-axis piezoelectric rotating stage equipped with resistive positional readout, we can control the orientation of the device relative to the applied magnetic field. One axis of the stage controls the tilt angle $\theta$ of the sample plane relative to the field (see Fig.~1a inset). The second axis allows for rotation of the sample stage about its normal, which controls the orientation of the in-plane component of the field relative to the sample (see the Supplemental Information for a more complete discussion). All measurements were performed with the same orientation of the in-plane field component unless otherwise noted. To calibrate the tilt angle $\theta$, we tuned the device to a regime where no anomalous Hall signal is present ($n = 0.04 n_s$) and used the Hall resistance $R_{yx}$ as a measure of the out-of-plane component of the field. We paid particular attention to precisely determining when the field was parallel to the sample plane (see the Supplemental Information). We define the tilt angle $\theta$ such that $90^\circ$ corresponds to a fully out-of-plane field while $0^\circ$ corresponds to a fully in-plane field as seen by the device. The in-plane component and out-of-plane component of the field are then $B_\parallel=B\cos(\theta)$ and $B_\perp=B\sin(\theta)$, respectively, where $B$ is the magnitude of the applied field.

We have measured hysteresis loops of $R_{yx}$ for different tilt angles of the device, with the device tuned to be ferromagnetic ($n/n_s = 0.746$ and $D/\epsilon_0 = -0.30\ \mathrm{V/nm}$). The coercive field, identified by the field at which the largest step in the Hall signal occurs, increases as the sample is rotated so that the field is in the plane of the sample (Fig.~1a).  When the hysteresis loops at the various angles are plotted as a function of the out-of-plane component of the magnetic field, $B_{\perp}$, the largest step in $R_{yx}$ consistently occurs at $119\pm1\ \mathrm{mT}$ for measured tilt angles down to about $9^\circ$ (Fig.~1b), indicating that the magnetization is indeed highly anisotropic and likely dominated by the orbital magnetic moment. 

The increased magnitude of the in-plane field as we lower the tilt angle does not strongly affect the hysteresis loops down to a loop performed at a tilt angle of $4.71^\circ \pm 0.10^\circ$ (plotted vs. $B_{\perp}$ in Fig.~1b and vs. $B$ in Fig.~2a): at this angle, a transition in $R_{yx}$ is still seen when out of plane field roughly matches the coercive field measured at larger tilt angles (marked with dashed vertical lines). At this tilt angle, the in-plane field reaches a maximum of $B_\parallel = 0.69\ \mathrm{T}$. However, as the sample is tilted closer to perfectly in-plane ($1.82^\circ \pm 0.10^\circ$ in Fig.~2b and beyond), we no longer see a dominant transition in $R_{yx}$ at the same value of out-of-plane field, and the measured magnitude of hysteresis in Hall resistance is significantly reduced. As the tilt angle is further reduced such that the out-of-plane field just reaches the coercive field (Fig.~2c) or does not reach the coercive field (Fig.~2d), any semblance of the hysteresis loops seen at larger tilt angles is lost. 

To explore whether the effects seen when the behavior in nearly in-plane field results from a small residual out-of-plane field, we compare hysteresis loops performed at small angles of similar magnitude but opposite sign (Fig.~2d). The two $R_{yx}$ vs. $B$ curves are very similar despite magnetic field angle deviating from in-plane in opposite directions, so the out-of-plane components are opposite for the two curves (see Fig.~S\ref{fig:fig2_xx_loops}d in the Supplemental Information for corresponding $R_{xx}$ curves). It is unlikely that this behavior results from a significant $B_{\perp}$: at $8\ \mathrm{T}$, the maximum $B$ applied in these loops, a tilt angle of more than $700\ \mathrm{mdeg}$ would be needed for $B_{\perp}$ to exceed the out of plane coercive field of $119\ \mathrm{mT}$. Such misalignment is well outside our experimental error for the two traces in Fig.~2D (the traces were performed at $+0.223^\circ \pm 0.049^\circ$ and $-0.171^\circ \pm 0.025^\circ$) so that, unlike in the other traces in Fig.~2, the out-of-plane coercive field is not reached (see the Supplemental Information for characterization of the rotating probe). Furthermore, were significant misalignment the source of the residual Hall signal, we would expect $R_{yx}$ to be antisymmetric in field and the device would likely recover some portion of its initial magnetization upon cycling the field, as is the case for larger tilt angles. For $1.82^\circ$ and $0.85^\circ$ the in-plane field has a significant effect: the shape of the hysteresis loop is quite different and the difference in $R_{yx}$ between the upward and downward sweeping traces remains substantial but approximately half that seen in a perpendicular field. The two traces in Fig.~2d happened to be acquired at different temperatures -- in perpendicular field we have found that the coercive field is $\sim 20\%$ less for the higher temperature while the size of the Hall signal is substantially unchanged~\cite{Sharpe2019}. A comparison performed at a constant temperature with the sample at a different in-plane angle yields similar results (Fig.~S\ref{fig:flip_oop2} of the Supplemental Information).

Up to now, hysteresis loops were acquired sequentially without explicitly repolarizing the sample between traces. To study the response of the orbitally polarized state to an in-plane field, we now start with the sample initially magnetized by an out-of-plane magnetic field. This training field is returned to zero, and the sample is then rotated to as close to in-plane as possible in zero magnetic field (for this measurement, the resultant tilt angle is $-57 \pm 21\ \mathrm{mdeg}$). Once rotated, a  magnetic field is applied at this very small angle to the sample (red trace in Fig.~3). As the in-plane field is increased, $|R_{yx}|$ initially rises, then begins to decrease at $B_\parallel=2.5\ \mathrm{T}$ (red trace in Fig.~3). As $B_\parallel$ is increased through $~5\ \mathrm{T}$, $|R_{yx}|$ rapidly falls, and we observe no hysteresis or steps in the Hall resistance for fields of larger magnitude, indicating an apparent transition from the Chern insulating state to a different state: above $~5\ \mathrm{T}$, both $R_{xx}$ and $R_{yx}$ show repeatable oscillations of order $2\ \mathrm{k\Omega}$ in size that appear to depend only on the magnitude of $B_\parallel$ (see the Supplemental Information). The accessible range of field is insufficient to say whether or not these oscillations are periodic, and if so whether the oscillations are periodic in $B_\parallel$ or $1/B_\parallel$. The high field state may or may not be polarized in spin and/or orbit. If it does have an orbital polarization, it is no longer set by the out-of-plane field component. The high-in-plane-field state also does not show a strong anomalous Hall signal. Were the Hall signal arising primarily from coupling to the magnetization, one would expect $R_{yx}(-M) = -R_{yx}(M)$, where $M$ is the magnetization of the sample. If high field were indeed fixing the orbital polarization, its direction and thus the Hall component of the signal should reverse with field direction. Instead $R_{yx}$ measured in this regime depends almost entirely on $|B|$ (Supplemental Information), so either this high-in-plane-field state lacks orbital polarization or the large in-plane field changes the topological character of bands so there is no net Chern number. Below $5\ \mathrm{T}$, the orbital polarization is being modified by in-plane field somehow, not by the small out of plane component. Upon decreasing $B_\parallel$, $R_{yx}$ shows evidence of magnetism but never recovers its initial value which nominally corresponds to a maximally polarized state (blue traces in Fig.~3). This behavior below $5\ \mathrm{T}$ may result from a repeatable pattern of out-of-plane orbital domains that is set by an in-plane field.

%%%%%%%%%%%%%%%%%%%%%%%%%%%%%%%%%%%%%%%%%%%%%%%%%%%%%%%%%%%%%%%%%%%%%
%$ Discussion and conclusions
%%%%%%%%%%%%%%%%%%%%%%%%%%%%%%%%%%%%%%%%%%%%%%%%%%%%%%%%%%%%%%%%%%%%%

Though this device does not exhibit quantized Hall resistance or zero longitudinal resistance, as previously noted it does appear to exhibit incipient Chern insulating behavior~\cite{Sharpe2019}. The simplest model for a Chern insulator at $3/4$ filling would have complete spin- and valley- polarization~\cite{Bultinck2020, Zhang2019b, Liu2021}. Other possible Chern insulator states have been considered in our previous publication~\cite{Sharpe2019}. 

As spin-orbit coupling is extremely weak in graphene, there should not be significant anisotropy in the direction spins prefer. In fact, there might be no relation between the direction of spin polarization and valley polarization. Starting from the out-of-plane orbitally polarized state with no external magnetic field, and then applying in-plane field, the initial rise in $|R_{yx}|$ (red trace in Fig.~3, up to 2T), may indicate that the spin is oriented by the external field, widening the gap to charge-carrying spin excitations which were causing departure from quantized transport. As $B_\parallel$ is increased beyond $~5\ \mathrm{T}$, we observe that $|R_{yx}|$ rapidly falls and is no longer hysteretic, perhaps indicating a field induced transition from the Chern insulator to some other state. If the low-field Chern insulator is valley-polarized but spin-unpolarized, polarizing spin by applying a large in-plane field could suppress the Hall signal by mixing with higher-order bands or by favoring spin instead of valley polarization~\cite{Kang2019, Zhang2019b}. Another possible mechanism for the observed transition is that, because of the finite thickness of tBLG, an in-plane field directly couples to the orbital moments~\cite{Lee2019} and a sufficiently large in-plane field could then drive the sample into a valley unpolarized state. Regardless of the dominant mechanism by which in-plane field couples to the device, it seems that either the bands are losing their topological character by the mixing in of higher bands~\cite{Kwan2019}, or the in-plane field is shifting population among a fixed set of flat bands. Thus far, in tBLG  aligned with hBN no evidence of magnetism has been observed at $n/n_s=1/4$~\cite{Sharpe2019, Serlin2020}, which should nominally be similar to $n/n_s=3/4$. Therefore there may be a competing state which does not have a net Chern number and would not exhibit a large Hall signal. Calculations show that both a gapless $C_2 T$-symmetric nematic state and a gapped $C_2 T$-symmetric stripe state are nearby in energy to the spin- and valley-polarized Chern insulator with $C=1$~\cite{Kang2020, BBChen2020, FXie2020, Pierce2021}. Evidence of such broken translational symmetry has recently been observed in tBLG as a series of Chern insulators which are inconsistent with the conventionally assigned moir\'e minibands but can be understood by doubling the unit cell~\cite{Pierce2021}.

We have observed that the magnetization of tBLG is sensitive primarily to the out-of-plane component of the field, requiring a threshold coercive field to flip the magnetization. A single value of $B_\perp$ required to switch the measured anomalous Hall resistance is consistent with uniaxial magnetization. It is unlikely that this uniaxial behavior is related to the electron spin because of the extremely low spin-orbit coupling in graphene. Rather, it is likely that the $3/4$ state is an orbital ferromagnet. The confinement of circulating electron currents to the plane would provide the high degree of anisotropy observed.

\clearpage
%%%%%%%%%%%%%%%%%%%%%%%%%%%%%%%%%%%%%%%%%%%%%%%%%%%%%%%%%%%%%%%%%%%%%
\section{Figures}
%%%%%%%%%%%%%%%%%%%%%%%%%%%%%%%%%%%%%%%%%%%%%%%%%%%%%%%%%%%%%%%%%%%%%
\begin{figure}
    \centering
    \includegraphics[width=450pt]{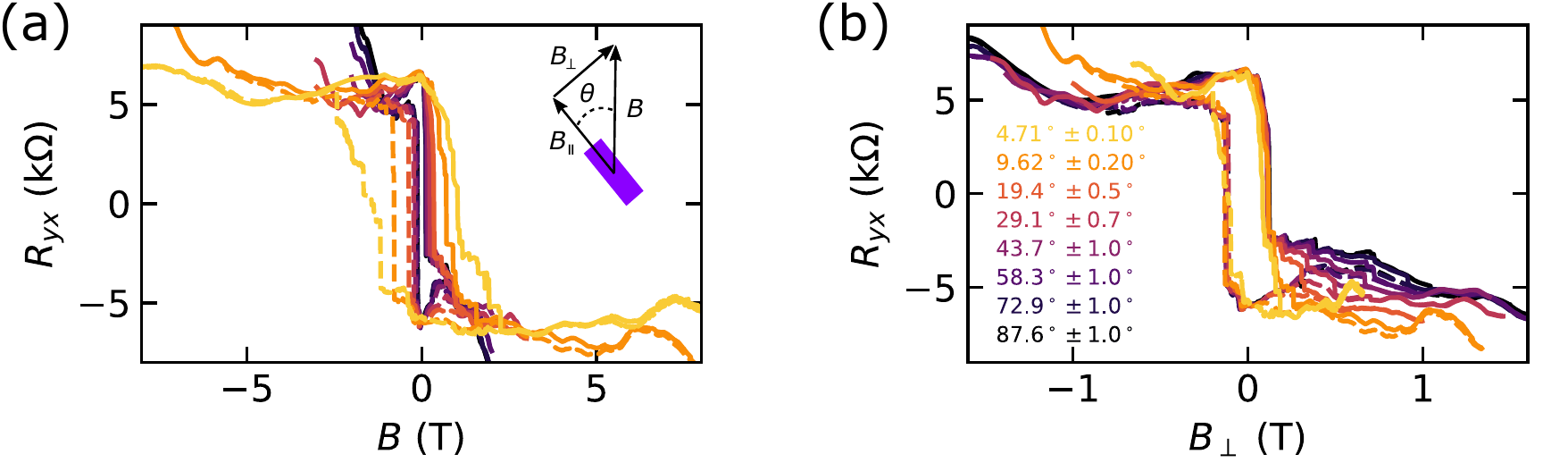}
    \caption{\textbf{Angular dependence of magnetic hysteresis loops.} Magnetic field dependence of the Hall resistance $R_{yx}$ with $n/n_s = 0.746$ and $D/\epsilon_0 = -0.30\ \mathrm{V/nm}$ at $29\ \mathrm{mK}$ as a function of the angle of the device relative to the field direction; $0^\circ$ corresponds to field in the plane of the sample. The hysteresis loops are plotted as a function of (a) the applied field $B$ and (b) the component of the field perpendicular to the plane of the sample $B_\perp$.
    The solid and dashed lines correspond to sweeping the magnetic field $B$ up and down, respectively. Inset: schematic diagram displaying the components of the magnetic field $B$ at the sample (shown in purple) for a given tilt angle $\theta$.
    }
    \label{fig:fig1}
\end{figure}

\begin{figure}
    \centering
    \includegraphics[width=450pt]{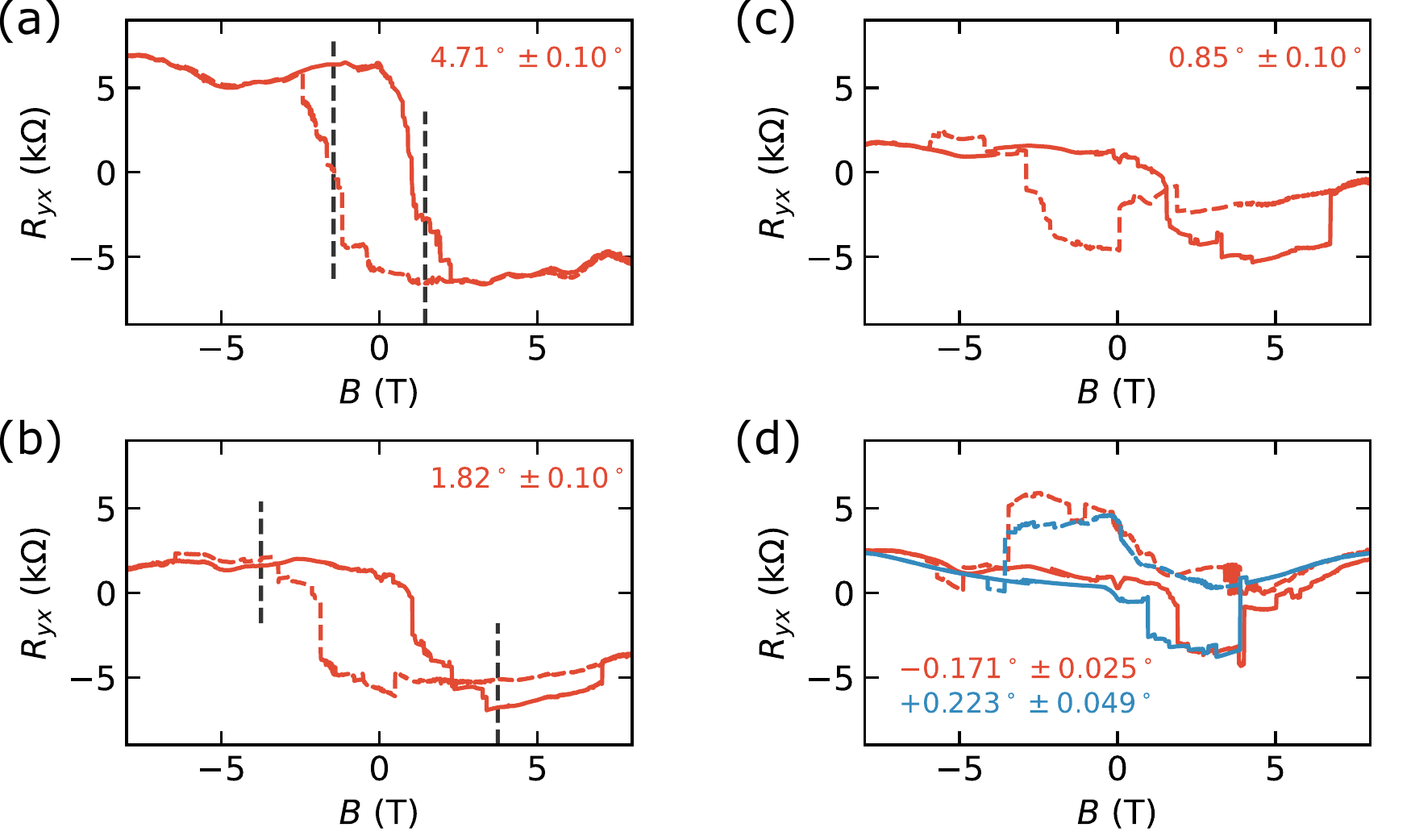}
    \caption{\textbf{Hysteresis loops for small tilt angles.}
    Angular dependence of $R_{yx}$ vs $B$ with $n/n_s = 0.746$ and $D/\epsilon_0 = -30\ \mathrm{V/nm}$ for angles of the field relative to the plane of the sample: (a) $4.71^\circ \pm 0.10^\circ$, (b) $1.82^\circ \pm 0.10^\circ$, (c) $0.85^\circ \pm 0.10^\circ$, (d) $+0.223^\circ \pm 0.049^\circ$, and $-0.171^\circ \pm 0.025^\circ$. Vertical dashed black lines indicate where the out-of-plane component of the field equals the coercive field $\pm 119\ \mathrm{mT}$. The out-of-plane component of the field is raised beyond the coercive field in panels (a),(b), just reaches the coercive field in (c), and does not reach it in (d). All traces were taken at $27\ \mathrm{mK}$ except for the trace with tilt angle $+0.223^\circ \pm 0.049^\circ$, which was taken at $1.35\ \mathrm{K}$.}
    \label{fig:fig2}
\end{figure}

\begin{figure}
    \centering
    \includegraphics[width=415pt]{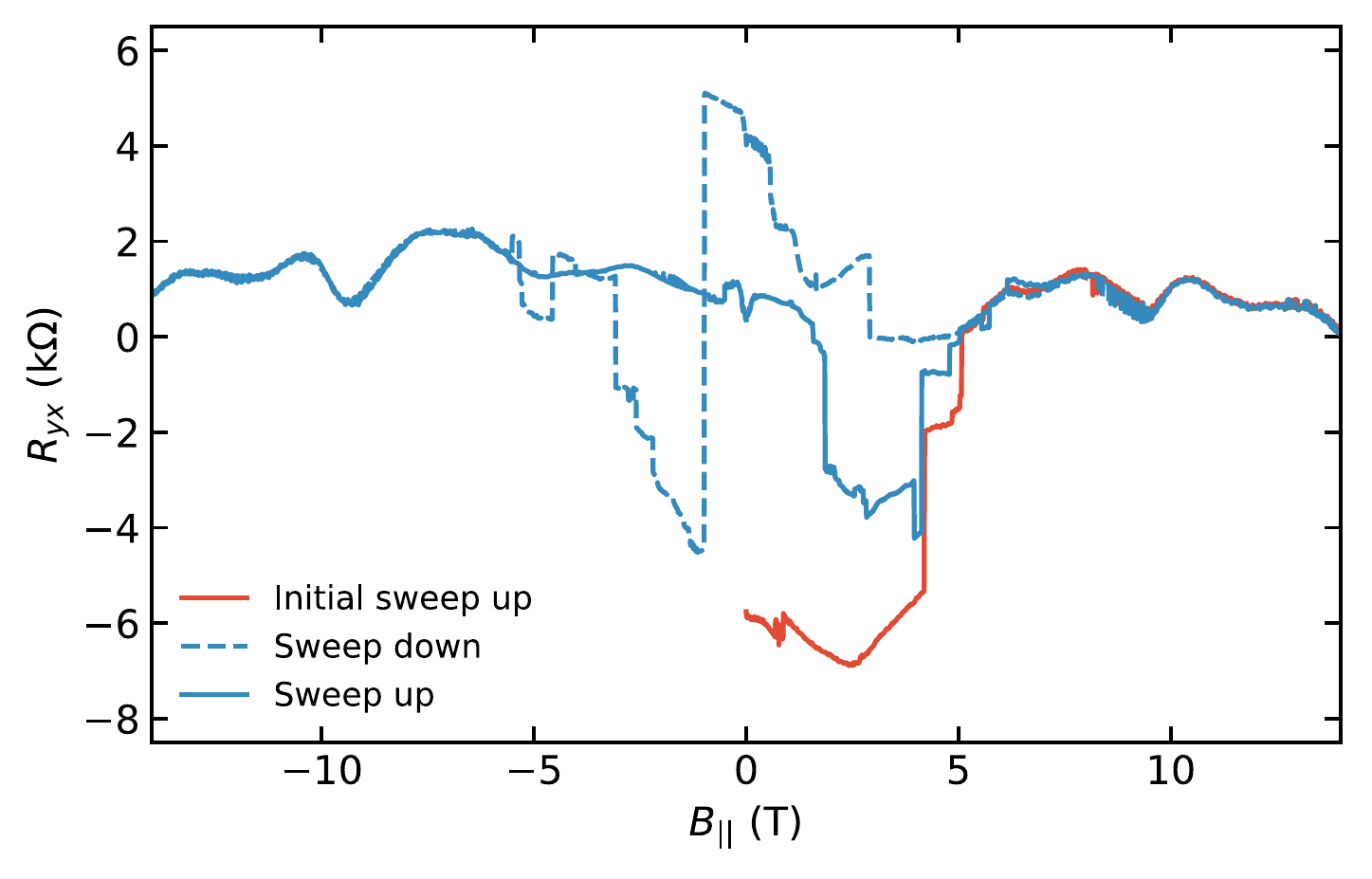}
    \caption{\textbf{Erasing the initial magnetic state.}
    In-plane hysteresis loops of $R_{yx}$ with $n/n_s = 0.746$ and $D/\epsilon_0 = -30\ \mathrm{V/nm}$ at $26\ \mathrm{mK}$. The sample is initially polarized with an out-of-plane field. The sample is then rotated to $-57 \pm 21\ \mathrm{mdeg}$ in zero magnetic field. The in-plane magnetic field $B_{\parallel}$ is then increased from zero (red trace) before completing a hysteresis loop (blue solid and dashed traces).}
    \label{fig:fig3}
\end{figure}

\clearpage

%%%%%%%%%%%%%%%%%%%%%%%%%%%%%%%%%%%%%%%%%%%%%%%%%%%%%%%%%%%%%%%%%%%%%
\begin{acknowledgement}
%%%%%%%%%%%%%%%%%%%%%%%%%%%%%%%%%%%%%%%%%%%%%%%%%%%%%%%%%%%%%%%%%%%%%
We acknowledge fruitful discussions with Greg Fuchs, Michael Zaletel, Allan MacDonald, T. Senthil, Ashvin Vishwanath, Eslam Khalaf, Oskar Vafek, Steve Kivelson, Yoni Schattner, Andrea Young, Matt Yankowitz, Feng Wang, and Guorui Chen. Some of these began at the Aspen Center for Physics, which is supported by National Science Foundation grant PHY-1607611. Yuan Cao and Pablo Jarillo-Herrero generously taught us about their fab process and their insights into tBLG. Hava Schwartz and Sungyeon Yang helped with device fabrication, and they and Anthony Chen performed preliminary measurements as part of a project-based lab class at Stanford. 
 
{\bf Funding:} Device fabrication, measurements, and analysis were supported by the U.S.\ Department of Energy, Office of Science, Basic Energy Sciences, Materials Sciences and Engineering Division, under Contract DE-AC02-76SF00515. Infrastructure and cryostat support were funded in part by the Gordon and Betty Moore Foundation through Grant GBMF3429. Part of this work was performed at the Stanford Nano Shared Facilities (SNSF), supported by the National Science Foundation under award ECCS-1542152. A.~S. acknowledges support from an ARCS Foundation Fellowship, a Ford Foundation Predoctoral Fellowship, and a National Science Foundation Graduate Research Fellowship. E.~F. acknowledges support from an ARCS Foundation Fellowship. K.W. and T.T. acknowledge support from the Elemental Strategy Initiative
conducted by the MEXT, Japan, Grant Number JPMXP0112101001, JSPS KAKENHI Grant Number JP20H00354 and the CREST (JPMJCR15F3), JST. {\bf Author contributions:} A.S. and J.F. fabricated devices. K.W. and T.T. provided the hBN crystals used for fabrication. A.S. and E.F. performed transport measurements. A.S., E.F., A.B., and J.F. analyzed the data. M.K. and D.G.-G. supervised the experiments and analysis. The manuscript was prepared by A.S. and E.F. with input from all authors.
{\bf Competing interests:} The authors have filed a patent disclosure on using the low-power switching of the magnetic state for low-temperature memory applications. M.K. was a member of the Science Advisory Board of the Gordon and Betty Moore Foundation until December 2019 and remains a chair of the DOE Basic Energy Science Advisory Committee. Both the Moore Foundation and Basic Energy Sciences provided funding for this work.
{\bf Data and materials availability:} The data from this study are available at the Stanford Digital Repository \cite{data_repo}.

\end{acknowledgement}

%%%%%%%%%%%%%%%%%%%%%%%%%%%%%%%%%%%%%%%%%%%%%%%%%%%%%%%%%%%%%%%%%%%%%
%% The same is true for Supporting Information, which should use the
%% suppinfo environment.
%%%%%%%%%%%%%%%%%%%%%%%%%%%%%%%%%%%%%%%%%%%%%%%%%%%%%%%%%%%%%%%%%%%%%
\begin{suppinfo}
\setcounter{figure}{0}    
\makeatletter
\renewcommand{\fnum@figure}{\figurename~S\thefigure}
\makeatother

\subsection{Methods}

Our device was previously characterized in a paper by the authors~\cite{Sharpe2019}. The device consists of twisted bilayer graphene (tBLG) encapsulated in two hexagonal boron nitride (hBN) cladding layers, each ${\sim} 50\ \mathrm{nm}$ thick. The heterostructure was assembled using a ``tear-and-stack'' technique~\cite{Cao2016, Kim2016}. A poly(bisphenol A carbonate) film/gel (Gel-Pak DGL-17-X8) stamp on a glass slide heated to $60\ ^\circ\mathrm{C}$ was used to pick up the top hBN flake. To stack two layers of graphene with a well defined twist angle, we used the Van der Waals attraction between hBN and monolayer graphene to tear off and pick up a portion of monolayer graphene from a larger flake. The remaining portion of monolayer graphene was then controllably rotated and picked up. The completed stack was transferred onto a 5x5 mm chip of $300$-$\mathrm{nm}$-thick SiO$_2$ atop degenerately doped Si substrate. The doped Si is used as a back gate.

The completed heterostructure was then fabricated into a measurable device using standard nanopatterning techniques. Patterned Ti/Au was deposited to serve as a top gate, and was then used as a hard mask for a  CHF$_3$/O$_2$ (50/5 sccm) etch to define a Hall bar geometry. During this etch, regions of the heterostructure were protected by resist extending outward from the top gate near each of the leads of the Hall bar to provide space for making Cr/Au edge contacts~\cite{LWang2013} without risk of shorting to the top gate. The sample temperature was kept below $180\ ^\circ\mathrm{C}$ throughout all processing in an effort to prevent relaxation of the twist angle of the tBLG.

The Au top gate and Si back gate can be used to tune both the carrier density in the tBLG and the displacement field applied to the device. The gates can be modeled as parallel plate capacitors such that the density under the top gated region is given by 
\[
n = C_{\text{BG}}(V_{\text{BG}}-V_{\text{BG}}^0) + C_{\text{TG}}(V_{\text{TG}}-V_{\text{TG}}^0),
\]
where $\mathrm{\text{BG}}$ ($\mathrm{\text{TG}}$) indicates the back (top) gate, $C$ is the capacitance per unit area determined from low-density Hall slope measurements, and $V_{\text{BG}}^0(V_{\text{TG}}^0)$ is the charge neutrality point of the back (top) gated region at zero displacement field. We define the applied displacement field as
\[
D = (D_{\text{BG}} - D_{\text{TG}})/2,
\]
where the displacement field within a given dielectric $D_i = \epsilon_i (V_i - V_i^0)/d_i$, $\epsilon_i$ is the relative dielectric constant, $d_i$ is the thickness of each dielectric, and $i=\{\text{BG}, \text{TG}\}$. The relative dielectric constant of hBN is assumed to be $\epsilon_{\text{TG}}=3$. As described previously~\cite{Sharpe2019}, we do not see any clear features to ascribe to a true zero in displacement field, so we assume that that the displacement field $D\approx 0$ when both gates are tuned to $0\ \mathrm{V}$. This assumption is reasonable given that the expected displacement field due to differences in the work functions between the top and back gate is small ($-0.01\ \mathrm{V/nm}$). Any nonzero displacement field when the gate voltages are zero should then simply yield a constant offset to our reported values. 

We mounted the sample in a Kyocera custom 32 contact ceramic leadless chip carrier (drawing PB-44567-Mod with no nickel sticking layer under gold, to reduce magnetic effects). The device was measured in a dilution refrigerator capable of reaching a base temperature of $30\ \mathrm{mK}$. The measurement lines are equipped with electronic filtering at the mixing chamber stage to obtain a low electron temperature in the device and reduce high-frequency noise. There are two stages of filtering. The wires are passed through a cured mixture of epoxy and bronze powder to filter GHz frequencies, then low-pass RC filters mounted on sapphire plates filter MHz frequencies. The sample was mounted in an attocube atto3DR two-axis piezoelectric rotating stage equipped with resistive positional readout.

Stanford Research Systems SR830 lock-in amplifiers with NF Corporation LI-75A voltage preamplfiers were used to perform four-terminal resistance measurements. A $1\ \mathrm{G\Omega}$ bias resistor was used to apply an AC bias current of $5\ \mathrm{nA}$ RMS at a frequency of $3.3373\ \mathrm{Hz}$. Keithley 2400 SourceMeters were used to apply voltages to the gates. All standard Hall configuration measurements were performed using the same voltage probes. One voltage contact behaved inconsistently and was not used in any of the reported measurements.

\subsection{Ascribing relative twist angles}

Analysis of an optical microscopy image of the completed heterostructure (Fig.~S\ref{fig:flake_afm} of Ref.~\citenum{Sharpe2019}) yields that the top graphene layer is rotated clockwise from the hBN by $1.0^\circ \pm 0.3^\circ$. Similar analysis of an atomic force microscopy image (Fig.~S\ref{fig:flake_afm}b) shows that the bottom graphene layer is rotated clockwise by $1.2^\circ \pm 0.4^\circ$ relative to the top graphene layer. Edges corresponding to specific layers of the tBLG are identified by comparing the atomic force microscopy image to an optical microscopy image of the graphene flake before it was torn (Fig.~S\ref{fig:flake_afm}a).

\begin{figure}
    \centering
    \includegraphics[width=450pt]{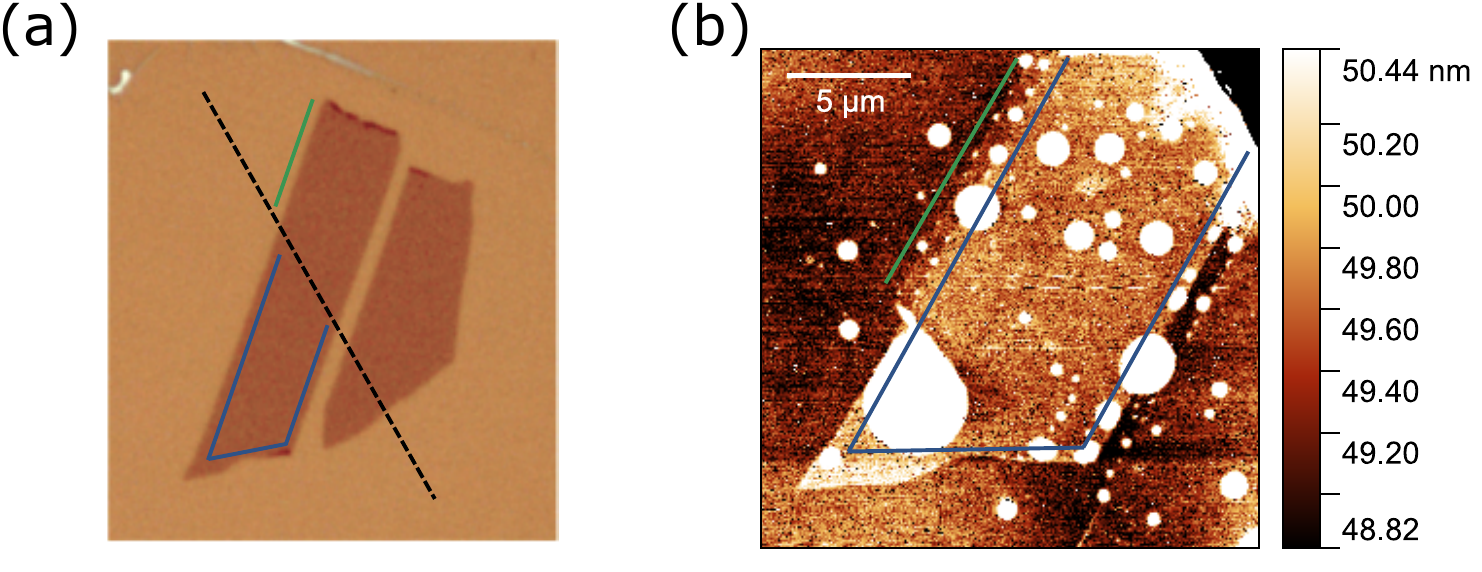}
    \caption{\textbf{Graphene flake prior to and after tearing.} 
    (a) Optical image of the graphene flake used in the complete device prior to tearing. The dashed black line indicates the approximate axis the flake was torn about. Blue lines indicate edges of the portion of the flake that will become the top layer of the tBLG. The green line indicates an edge on the portion of the flake that will become the bottom layer of the tBLG. The second graphene flake to the right of the annotated flake was not used to fabricate a device. The grey lines above the graphene flakes are tape residue.
    (b) Atomic force microscopy image of the completed heterostructure prior to device fabrication. Specific edges are indicated in the same manner as they were in (a). The green line has been manually rotated by $1.2^\circ$ clockwise relative to the rightmost blue line. The device was fabricated in a region without any bubbles.}
    \label{fig:flake_afm}
\end{figure}

\subsection{Rotator calibration}

\subsubsection*{Summary:}

To control the orientation of the device relative to the field from the solenoid, the sample chip carrier was mounted in an attocube atto3DR two-axis piezoelectric rotator. 28 of the 32 contacts on the Kyocera chip carrier were available for measurement given space constraints in the probe. The atto3DR combines two rotators: one to control the tilt angle $\theta$ of the sample relative to the field to tune the relative magnitudes of the out-of-plane and in-plane components of the field (see Fig.~1a inset of the main text) and one to rotate about the normal of the stage to control the direction of the in-plane component of the field relative to the sample. Rotation is eucentric: the sample location in the solenoid is fixed as its orientation is changed. The rotator is equipped with a resistive readout for each axis for determining the angular position of the stage. We calibrated this readout using the Hall resistance $R_{yx}$ of the device in a regime where the (ordinary) Hall effect is a measure of the out-of-plane field. All the data we present are accompanied by estimates of the angular position and error bounds based on this calibration. A precise calibration is particularly important when the field is nearly in-plane to avoid a significant unintended out-of-plane component when the magnitude of the total field is large. For all our nominally in-plane field measurements, the out-of-plane component remained smaller than the out-of-plane coercivity.

\subsubsection*{Details:}

Since samples may not sit perfectly flat in the ceramic chip carriers we use to mount them, and the chip carriers may not always sit in exactly the same position in the pogo pin socket of the rotator stage, with every cooldown it is necessary to calibrate the angular position of a sample as a function of the resistive position readout for accurate positioning. To calibrate the tilt angle $\theta$ for this device, we tuned the device to a carrier density $n = 0.04 n_s$ where no anomalous Hall effect was present and a linear ordinary Hall effect allowed us to extract the out-of-plane component of the field. The resistive position readout was then calibrated by rotating the sample in a fixed field while measuring the Hall resistance and the resistive readout. The measured Hall resistance may have a small offset at zero applied field due to mixing in of the longitudinal resistivity. Therefore, to ensure an accurate identification of the angle corresponding to an in-plane field, we rotated the sample in both a $+6\ \mathrm{T}$ field and a $-6\ \mathrm{T}$ field. The in-plane-field position was then determined as the angle corresponding to the value of the resistive readout where the measured $R_{yx}$ from the two angular sweeps were equal.

Uncertainty in the true angular position arises from several sources. Noise in the measurement of the resistive position readout can be reduced, but not completely removed, by measuring with a lock-in amplifier. This noise, along with uncertainty in the relationship between the Hall signal used for calibration and the true out-of-plane field component, contributes to error in the calibration. The resistive readout is also hysteretic with rotation because of backlash in the piezoelectric rotators. This backlash can be controlled for by consistently rotating to a final position from the same direction or generating a separate calibration for each direction of rotation. Additionally, the sample will have some small tilt relative to the rotator stage. Thus, except when the normal to the rotator stage is aligned with the solenoid axis, rotation of the in-plane angle will change the tilt angle of the sample relative to the field. One solution to this problem is simply to calibrate $\theta$ for each in-plane angle used. Finally, a large magnetic field can apply a torque to the sample or rotator. We have observed a resulting field-induced rotation in a separate Bernal bilayer graphene sample under nearly in-plane field: compared to the angle calibrated in a field of $6\ \mathrm{T}$ magnitude, $\theta$ can change by up to $15\ \mathrm{mdeg}$ at zero field, and $40\ \mathrm{mdeg}$ at $14\ \mathrm{T}$. This rotation is measurable using a Hall signal, but is not reflected in the rotator resistive readout. The torque-induced rotation is equal rather than opposite for opposite magnetic field directions (i.e. opposite solenoid current). We account for all of these sources of error in our angular position uncertainty estimates, except for field-induced rotation. This omission is justified because our calibration is performed at 6 T, roughly the maximum field at which we observe sharp transitions in the Hall signal in nearly in-plane field; the provided error estimate therefore accurately characterizes the maximum possible out-of-plane field during these transitions.

Based on the precision and accuracy of the angular position calibration near $\theta = 0$ (in-plane field), we are confident that for all measurements in nominally in-plane field, the out-of-plane field component remained substantially smaller than the out-of-plane coercivity of the tBLG device, which was approximately $119\ \mathrm{mT}$ for the density and displacement field used. To achieve a $119\ \mathrm{mT}$ out-of-plane component in a $6\ \mathrm{T}$ field requires $|\theta| \ge 1.136^\circ$, whereas of our nominally in-plane sweeps the largest deviation from zero angle is ~200 mdeg; the magnetic transitions observed in the Hall signal between $4$ and $6\ \mathrm{T}$ in Figs.~2(d) and 3 of the main text therefore do not appear to be driven simply by the out-of-plane field component flipping an orbital magnetization of a domain. Even at $14\ \mathrm{T}$, the out-of-plane component remains below $119\ \mathrm{mT}$ for $|\theta| < 487$ mdeg, and each of the measurements shown in Figs.~2(d) and 3 of the main text and Figs.~S\ref{fig:fig2_xx_loops}D, S\ref{fig:fig3_xx}, S\ref{fig:fig3_asym}, S\ref{fig:fig3_abs}, S\ref{fig:flip_oop2}, and S\ref{fig:xx_inplane} of the supplement information are well within this range, even accounting for possible field-induced rotation of the rotator.

\subsection{Longitudinal resistance data}

In this section, we provide the corresponding longitudinal resistance data for the figures of the main text. Fig.~S\ref{fig:fig1_xx_loops} shows the longitudinal resistance $R_{xx}$ corresponding to the Hall resistance data of Fig.~1 of the main text. As was seen previously with this device~\cite{Sharpe2019}, $R_{xx}$ displays visible hysteresis, presumably due to mixing in of the Hall signal from inhomogeneity in the device or its domain structure. When plotted as a function of the perpendicular field component (Fig.~S\ref{fig:fig1_xx_loops}b), we see that the longitudinal resistance depends mostly on the perpendicular field component with some small variations from angle to angle, perhaps due to an effect of the in-plane magnetic field.

Fig.~S\ref{fig:fig2_xx_loops} shows the longitudinal resistance $R_{xx}$ corresponding to the Hall resistance data of Fig.~2 of the main text. As the angle of the sample is tuned closer and closer to a purely in-plane magnetic field, the variations in the longitudinal resistance become smaller. As the field becomes almost perfectly in-plane (Fig.~S\ref{fig:fig2_xx_loops}d), the magnitude of hysteretic $R_{yx}$ is diminished compared to that seen at larger out-of-plane tilt angles. We see jumps in the longitudinal resistance that likely correspond to the flipping of magnetic domains. 

Finally, Fig.~S\ref{fig:fig3_xx} shows the (a) longitudinal and (b) Hall response to a purely in-plane field of a state initially magnetized out-of-plane. Panel (b) is reproduced from Fig.~3 of the main text. As was discussed in the main text, applying an in-plane magnetic field erases the initial magnetic state and drives a phase transition to a different state above 5 T in-plane field. Below this critical field, the magnetic state is recovered but the full polarization of the system is not. This leads to a reduction in the magnitude of the Hall resistance. 

\begin{figure}
    \centering
    \includegraphics[width=450pt]{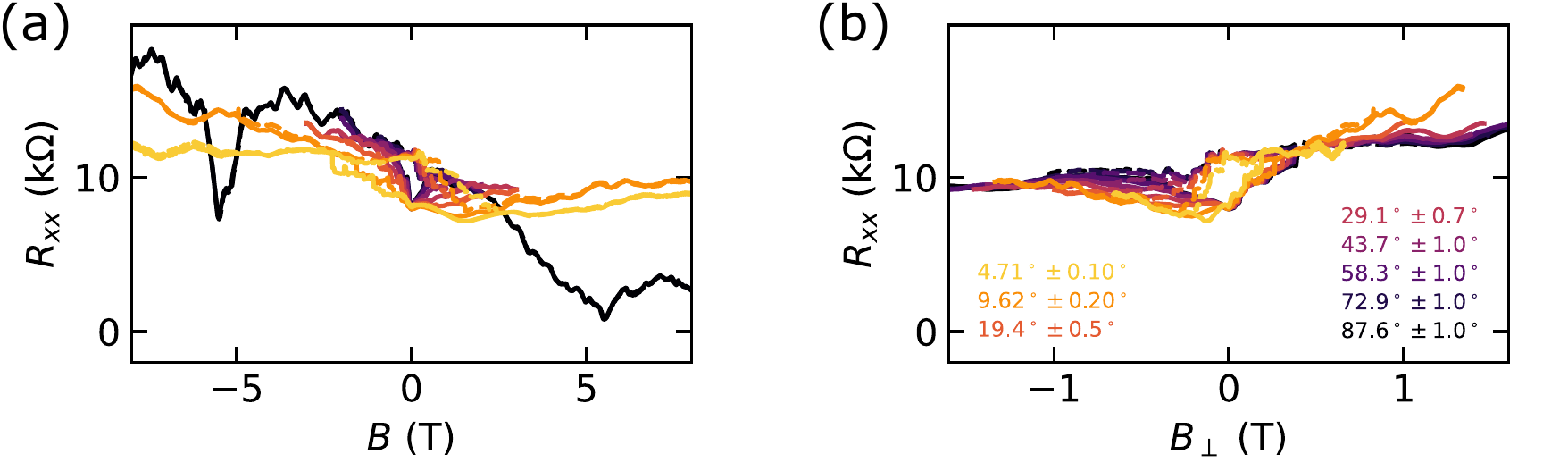}
    \caption{\textbf{Angular dependence of longitudinal resistance in magnetic hysteresis loops.} Magnetic field dependence of the longitudinal resistance $R_{xx}$ corresponding to the data shown in Fig.~1 of the main text, with $n/n_s = 0.746$ and $D/\epsilon_0 = -0.30\ \mathrm{V/nm}$ at $29\ \mathrm{mK}$ as a function of the angle of the device relative to the field direction; $0^\circ$ corresponds to field in the plane of the sample. The hysteresis loops are plotted as a function of (a) the applied field $B$ and (b) the component of the field perpendicular to the plane of the sample $B_\perp$.
    The solid and dashed lines correspond to sweeping the magnetic field $B$ up and down, respectively.}
    \label{fig:fig1_xx_loops}
\end{figure}

\begin{figure}
    \centering
    \includegraphics[width=450pt]{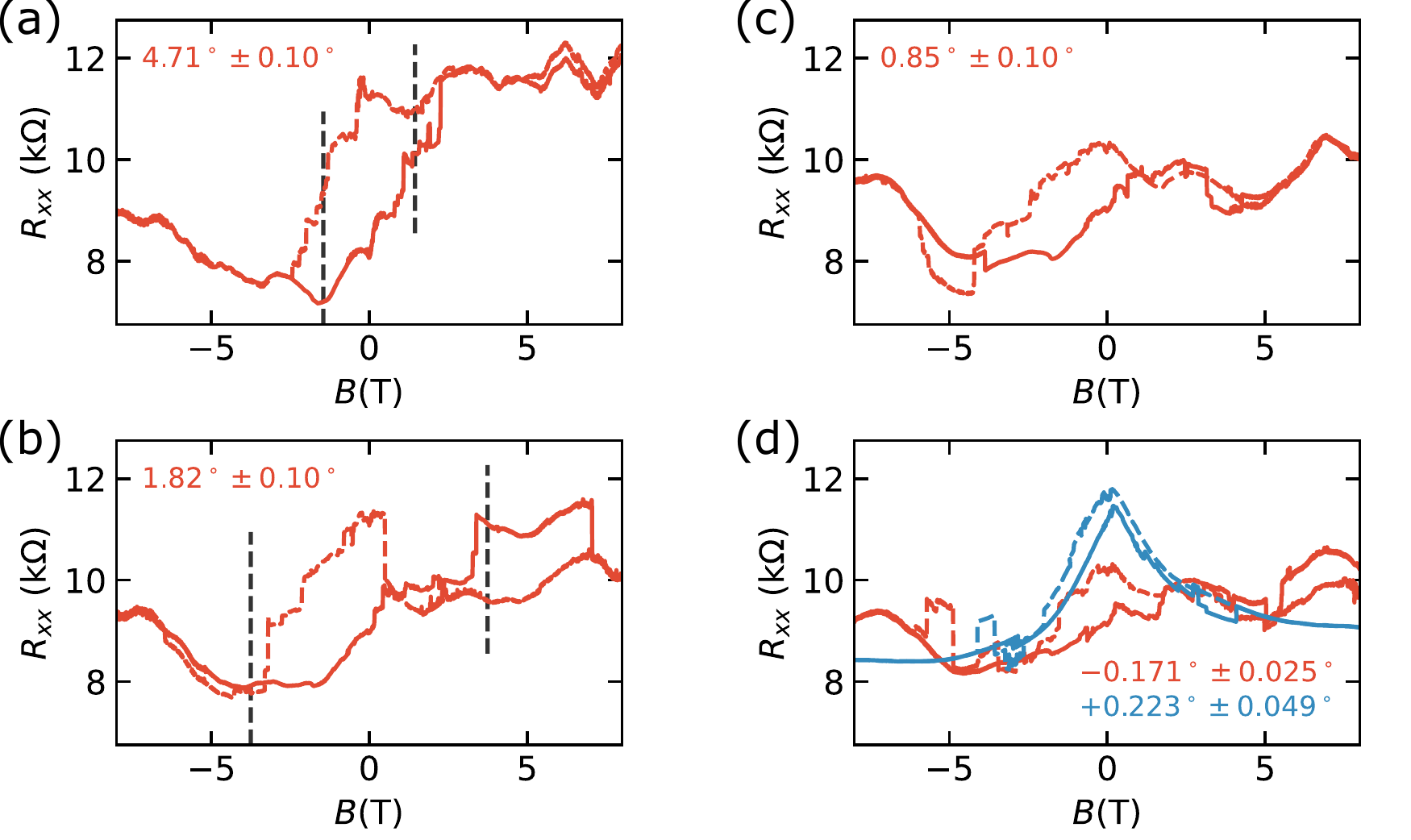}
    \caption{\textbf{Longitudinal resistance hysteresis loops for small tilt angles.} Angular dependence of the longitudinal resistance $R_{xx}$ vs $B$ corresponding to the data shown in Fig.~2 of the main text with $n/n_s = 0.746$ and $D/\epsilon_0 = -30\ \mathrm{V/nm}$ for angles of the field relative to the plane of the sample: (a) $4.71^\circ \pm 0.10^\circ$, (b) $1.82^\circ \pm 0.10^\circ$, (c) $0.85^\circ \pm 0.10^\circ$, (d) $+0.223^\circ \pm 0.049^\circ$, and $-0.171^\circ \pm 0.025^\circ$. Vertical dashed black lines indicate where the out-of-plane component of the field equals the coercive field $\pm 119\ \mathrm{mT}$.  The out-of-plane component of the field is raised beyond the coercive field in panels (a),(b), just reaches the coercive field in (c), and does not reach it in (d).All traces were taken at $27\ \mathrm{mK}$ except for the trace with tilt angle $+0.223^\circ \pm 0.049^\circ$, which was taken at $1.35\ \mathrm{K}$.}
    \label{fig:fig2_xx_loops}
\end{figure}

\begin{figure}
    \centering
    \includegraphics[width=450pt]{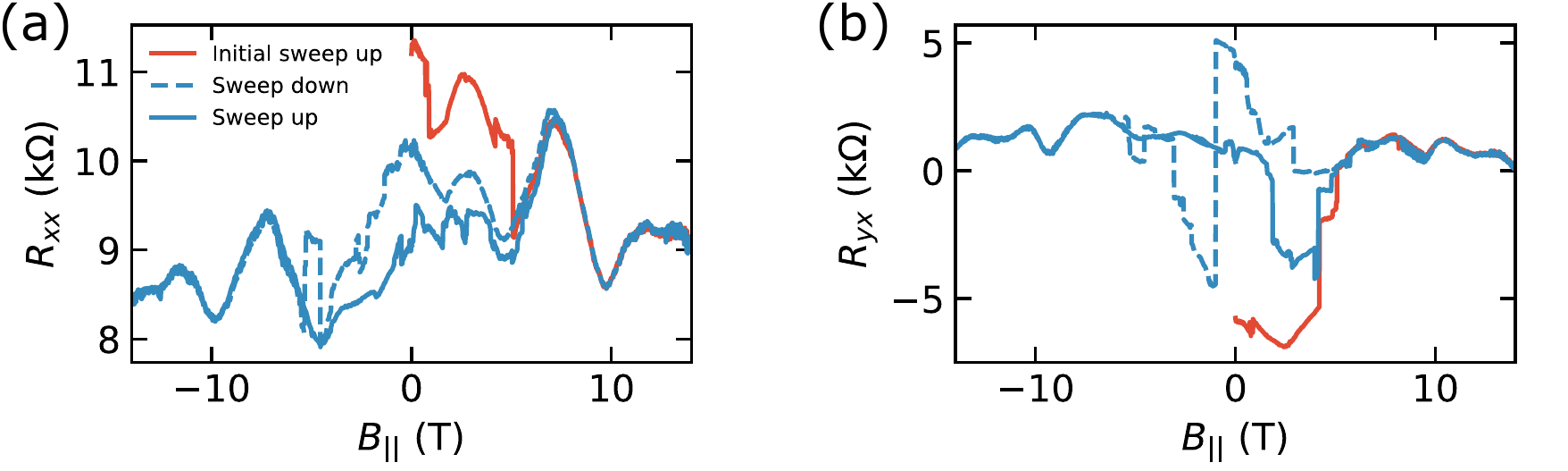}
    \caption{\textbf{Longitudinal resistance under an in-plane field from an initially magnetized state.}
    In-plane hysteresis loops of the (a) longitudinal resistance $R_{xx}$ corresponding to the data shown in Fig.~3 of the main text with $n/n_s = 0.746$ and $D/\epsilon_0 = -30\ \mathrm{V/nm}$ at $26\ \mathrm{mK}$. The Hall resistance data $R_{yx}$ of Fig.~3 of the main text are replicated in panel (b). The sample is initially polarized with an out-of-plane field. The sample is then rotated to $-57 \pm 21 \mathrm{mdeg}$ in zero magnetic field. The field $B_{\parallel}$ is then increased from zero (red trace) before completing a hysteresis loop (blue solid and dashed traces).}
    \label{fig:fig3_xx}
\end{figure}

\clearpage
\subsection{Antisymmetric components and dependence on magnitude of in-plane field}

We can antisymmetrize the hysteresis loop shown in Fig.~3 of the main text by considering $R_{yx}^{\mathrm{asym}}(B) = (R_{yx}(B,\, \mathrm{sweeping\ upwards}) - R_{yx}(-B,\, \mathrm{sweeping\ downwards}))/2$ (Fig.~S\ref{fig:fig3_asym}a). The final upward sweep is completed only up to $8\ \mathrm{T}$, so we have restricted the symmetrization to $[-8\ \mathrm{T},8 \mathrm{T}]$. We see that for small in-plane fields, the system is strongly hysteretic. Although a typical ferromagnetic hysteretic loop is not observed as a function of in-plane field, we do see that there is a large discrepancy between the up and down sweeps. Above an in-plane field of $\sim5\ \mathrm{T}$, the up and down sweeps  differ only slightly. Additionally, the magnitude of the antisymmetric component of the Hall signal is quite small, $\lesssim 0.5\ \mathrm{k\Omega}$, over this high field range, in contrast to the values at lower in-plane fields, or any out-of-plane fields. A comparison with a loop performed at a tilt angle of $87.6 ^\circ \pm 1.0^\circ$ such that the field is nearly out-of-plane (Fig.~S\ref{fig:fig3_asym}b) shows that this symmetrization process preserves the hysteresis loop (which is shown schematically in Fig.Fig.~S\ref{fig:fig3_asym}c).

\begin{figure}
    \centering
    \includegraphics[width=450pt]{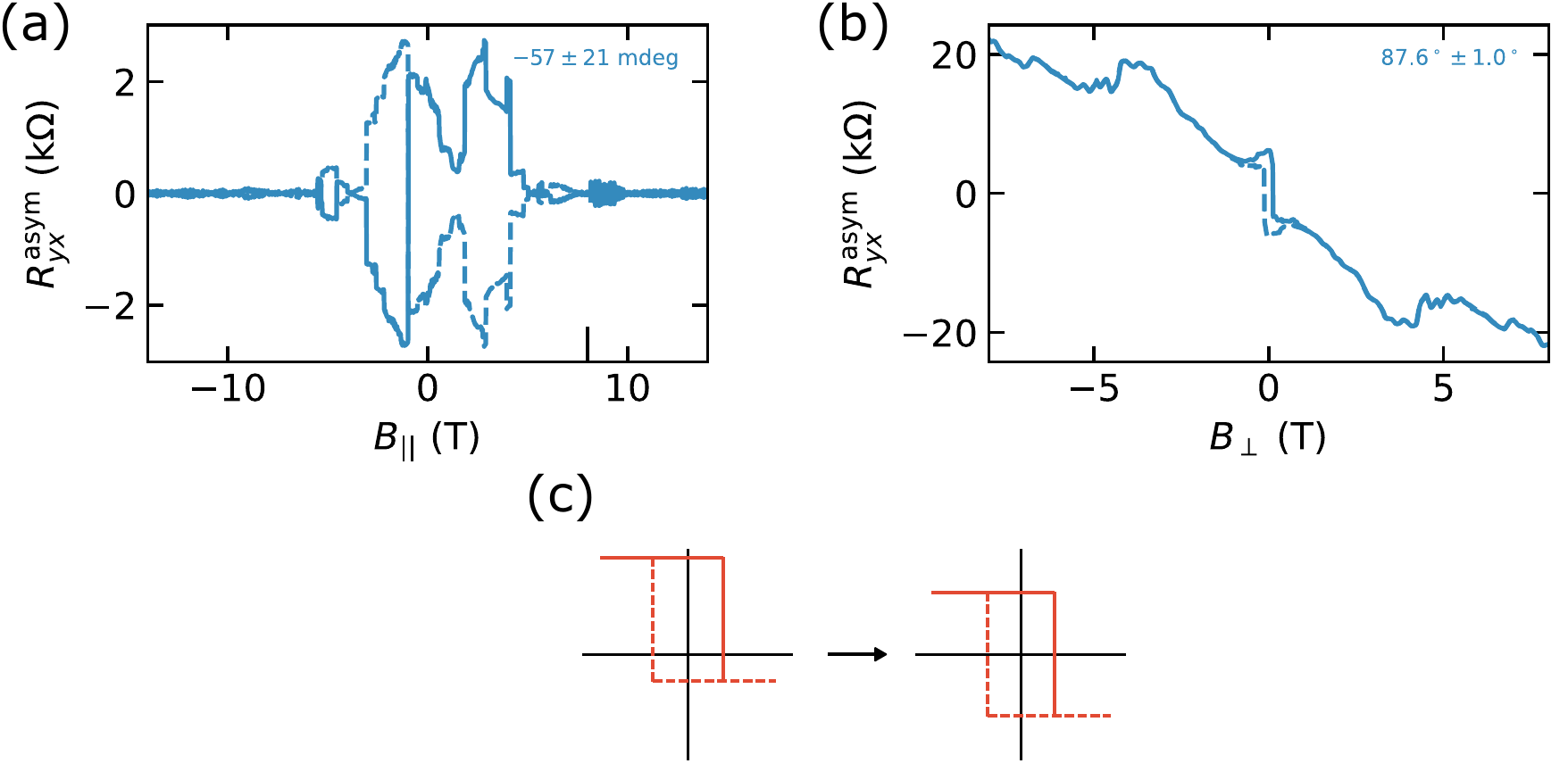}
    \caption{\textbf{Antisymmetric contribution of the Hall resistance under an in-plane field.}
    (a) Antisymmetric component of the Hall resistance $R_{yx}^{\mathrm{asym}}$ corresponding to the Hall data shown in Fig.~3 of the main text. $\pm R_{yx}^{\mathrm{asym}}$ is plotted as a solid (dashed) line. For $8\ \mathrm{T}$ and above (indicated by the large tick on the horizontal axis), we report $R_{yx}^{\mathrm{asym}}$ for the initial sweep up and the sweep down. Otherwise we report $R_{yx}^{\mathrm{asym}}$ for the sweep down and the sweep up (which was only completed up to $8\ \mathrm{T}$.) (b) Antisymmetric component of the Hall resistance $R_{yx}^{\mathrm{asym}}$  for the nearly out-of-plane hysteresis loop performed at $87.6^\circ \pm 1.0^\circ$, shown in Fig.~1 of the main text. (c) Schematic diagram of the symmetrization process for an ideal hysteresis loop that is offset from zero in the vertical direction.
    }
    \label{fig:fig3_asym}
\end{figure}

When the data from Fig.~S\ref{fig:fig3_xx} are plotted against the magnitude of the field (shown in Fig.~S\ref{fig:fig3_abs} for fields above 6 T), we again see that those parts of the data outside of $[-6\ \mathrm{T},6 \mathrm{T}]$ are remarkably similar. For large in-plane fields, the longitudinal and Hall resistances are qualitatively similar, and each shows an offset of order $0.5\ \mathrm{k\Omega}$ between data for the two field polarities. Expanding on what we said in the main text, the similarity of the longitudinal and Hall resistance and the fact that both are mostly symmetric in field suggest that the sample is no longer orbitally polarized and that the apparent Hall signal results from  mixing in of the longitudinal signal. 

\begin{figure}
    \centering
    \includegraphics[width=450pt]{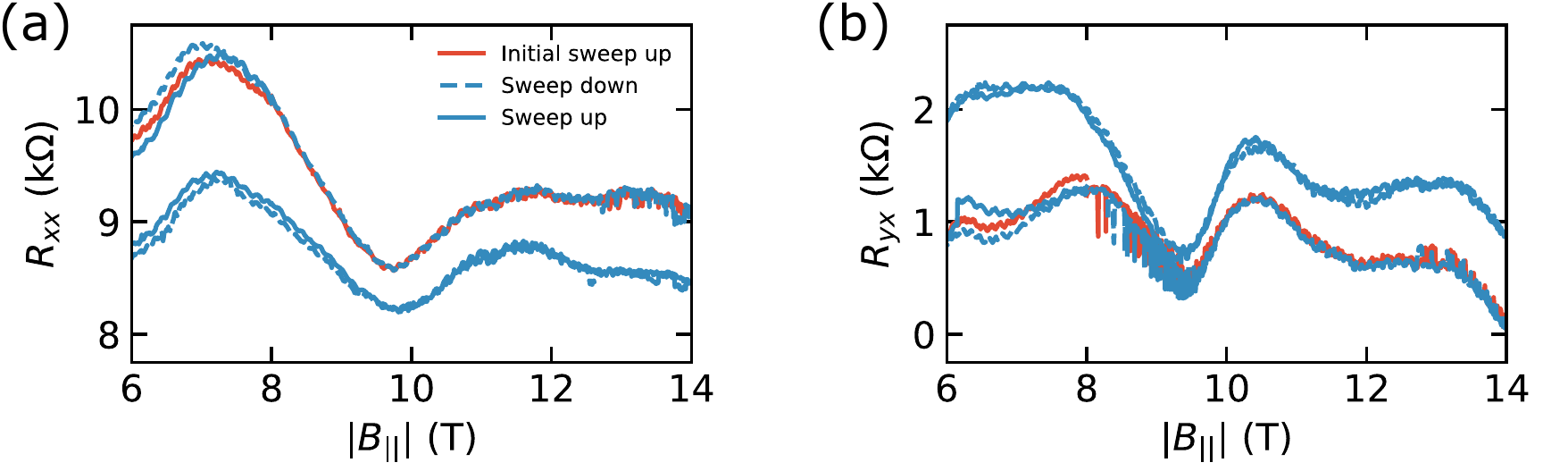}
    \caption{\textbf{Dependence on the magnitude of the in-plane field.}
    In-plane hysteresis loops of corresponding longitudinal resistance $R_{xx}$ to the data shown in Fig.~3 of the main text with $n/n_s = 0.746$ and $D/\epsilon_0 = -30\ \mathrm{V/nm}$ at $26\ \mathrm{mK}$. The sample is initially polarized with an out-of-plane field. The sample is then rotated to $-57 \pm 21\ \mathrm{mdeg}$ in zero magnetic field. The field $B_{\parallel}$ is then increased from zero (red trace) before completing a hysteresis loop (blue traces). Though the absolute resistance is offset between the two panels, as seen in the vertical axis labels, the size of the resistance range is the same in both panels.}
    \label{fig:fig3_abs}
\end{figure}

\subsection{Effect of out-of-plane field at small tilt angles}

In this section we present additional data similar to those in Fig.~2d to elucidate the effect of the sign of a small out-of-plane field in the background of a large in-plane field at small tilt angles (Fig.~S\ref{fig:flip_oop2}). Note that for the presented data, the sample has been rotated in the plane by $20^\circ$ relative to the traces performed in Fig.~2d so that the in-plane field is in a different direction relative to the sample's crystal axis and direction of current flow. Between the two traces shown in Fig.~S\ref{fig:flip_oop2}, there are numerous jumps in $R_{yx}$, some which occur at similar values of $B$ and some which are slightly shifted between the two traces. There is an order of magnitude difference in the magnitude of the out-of-plane field between the two traces, suggesting that the features in common between the two traces of Fig.~S\ref{fig:flip_oop2} are primarily a response to the in-plane field. 

\begin{figure}
    \centering
    \includegraphics[width=415pt]{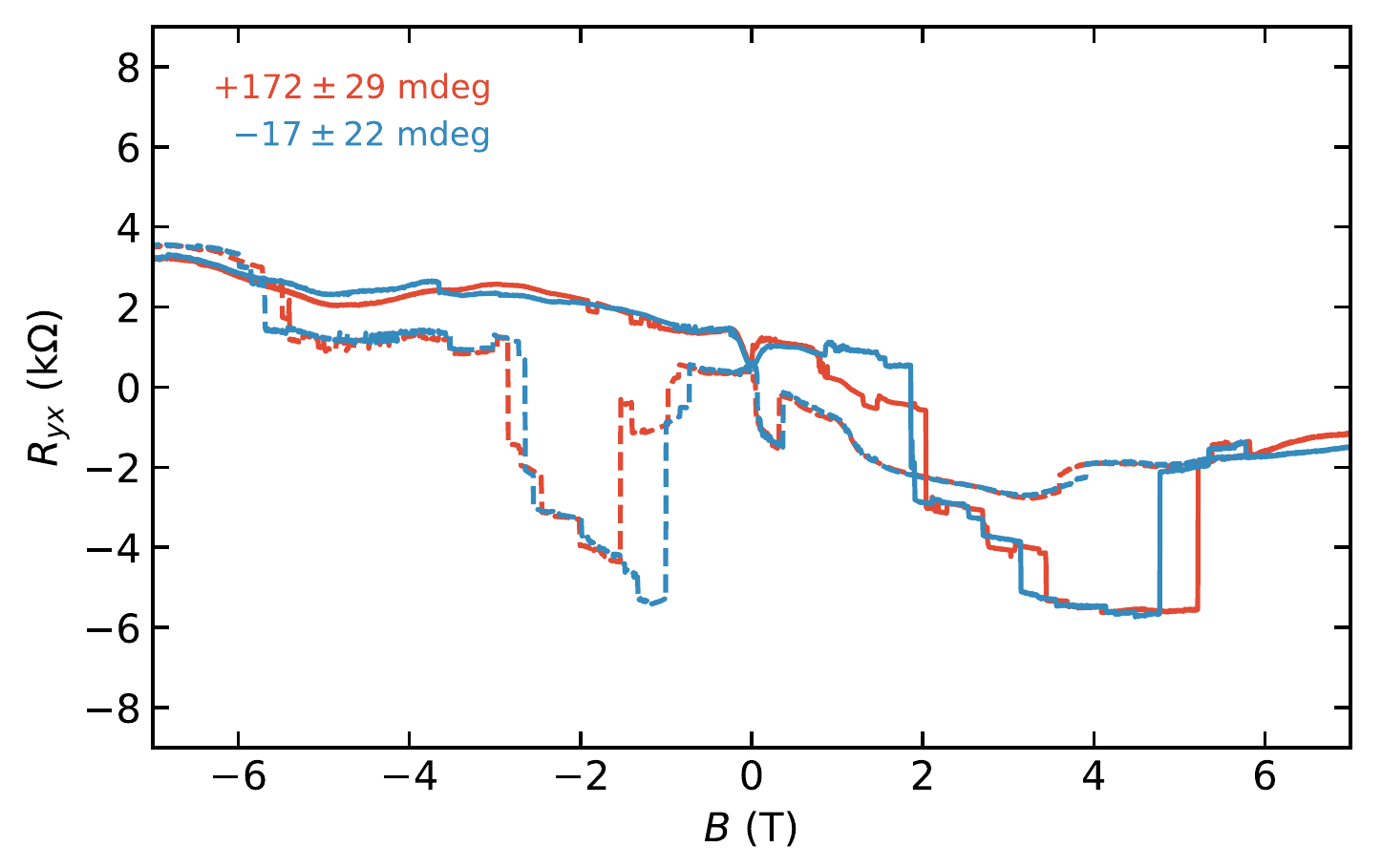}
    \caption{\textbf{Hysteresis loops at small angles.}
    Magnetic field hysteresis loops where $R_{yx}$ is measured at two small angles of very different magnitude and likely opposite sign: $+172 \pm 29\ \mathrm{mdeg}$ in red and $-17 \pm 22\ \mathrm{mdeg}$ in blue. Both traces were taken at $28\ \mathrm{mK}$ with $n/n_s = 0.746$ and $D/\epsilon_0 = -30\ \mathrm{V/nm}$. The sample has been rotated in the plane by an angle of $20^\circ$ relative to the measurements performed in Fig.~2 of the main text.}
    \label{fig:flip_oop2}
\end{figure}

%%%%%%%%%%%%%%%%%%%%%%%%%%%
\subsection{Effect of in-plane field on longitudinal resistance}

For completeness, we have examined the effect of an in-plane field at other carrier densities. The longitudinal resistance $R_{xx}$ does not depend significantly on $B_\parallel$ except in the range of densities $n/n_s = 0.1$ to $0.65$, where $R_{xx}$ increases with increasing $B_\parallel$ (Fig.~S\ref{fig:xx_inplane}). This is the opposite dependence compared to the behavior in a superconducting device presented in Ref.~\citenum{Yankowitz2019}.

\begin{figure}
    \centering
    \includegraphics[width=415pt]{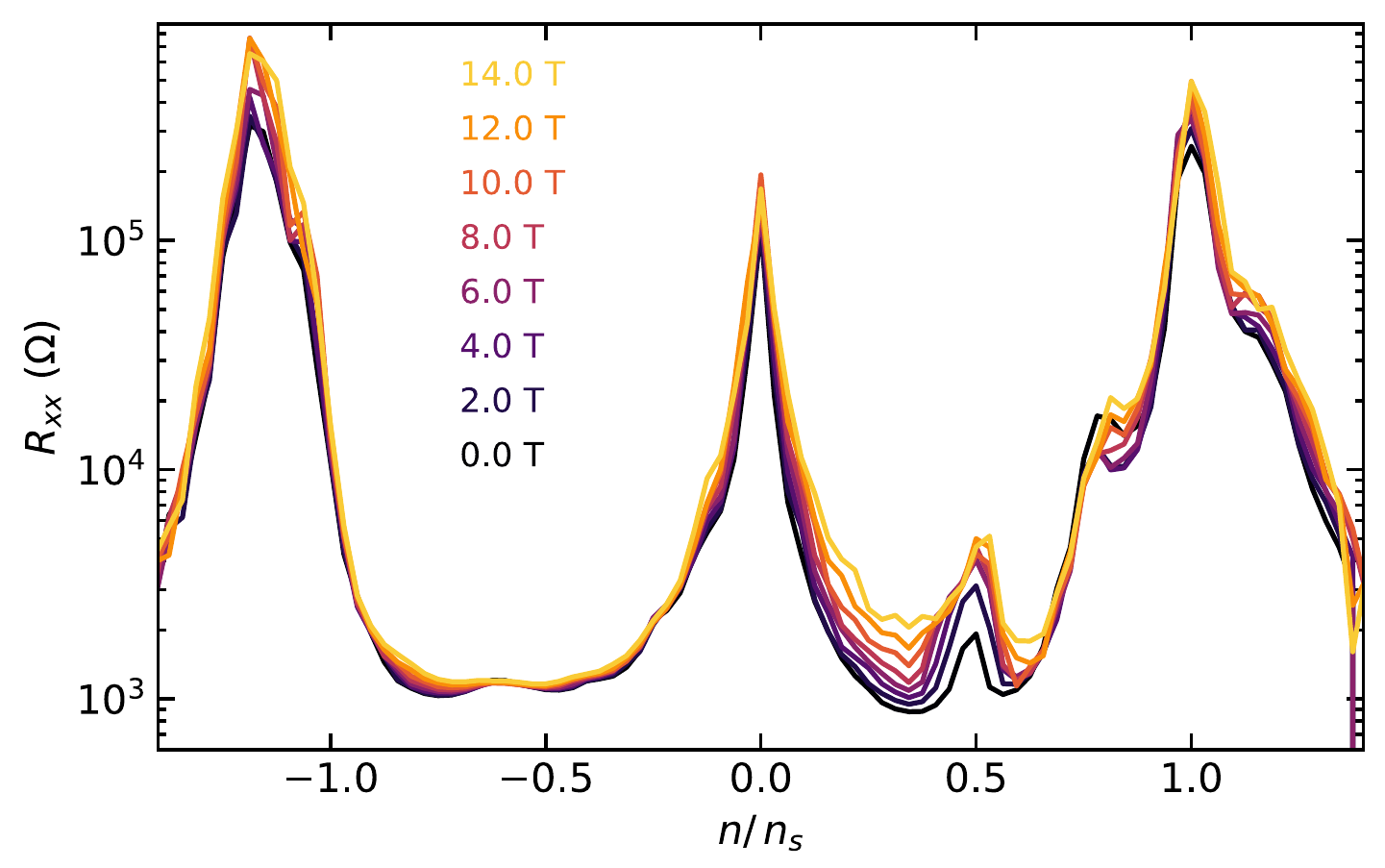}
    \caption{\textbf{In-plane field dependence of longitudinal resistance.}
    Longitudinal resistance $R_{xx}$ as a function of carrier density $n$ for several different in-plane magnetic fields at a fixed displacement field of $D/\epsilon_0 = -0.30\ \mathrm{V/nm}$ and $1.2\ \mathrm{K}$ for a tilt angle of $-62 \pm 23\ \mathrm{mdeg}$.}
    \label{fig:xx_inplane}
\end{figure}

\clearpage

\end{suppinfo}

\end{document}